\documentclass[aps,pre,twocolumn,floatfix]{revtex4-2}
\usepackage{amsmath,amssymb}
\usepackage{graphicx}
\newcommand{\Tr}{\textit{Tr}\,}
\newcommand{\tauav}[1]{\langle#1\rangle}
\newcommand{\DD}{\tauav{\hat{D}}}
\newcommand{\tDD}{\tauav{\tau\hat{D}}}

\usepackage{color}

\begin{document}

\title{The first detection time of a quantum state under random probing}

 \author{David A. Kessler}
 \email{kessler@dave.ph.biu.ac.il}
 \affiliation{Department of Physics,  Bar-Ilan University, Ramat-Gan 52900, Israel}
 
 \author{Eli Barkai}
 \email{Eli.Barkai@biu.ac.il}
 \affiliation{Department of Physics, Institute of Nanotechnology and Advanced Materials, Bar-Ilan University, Ramat-Gan 52900, Israel}

 \author{Klaus Ziegler}
\email{klaus.ziegler@physik.uni-augsburg.de}
 \affiliation{Institut f\"ur Physik, Universit\"at Augsburg, D-86135 Augsburg, Germany}

 \begin{abstract}
 We solve for the statistics of the first detection  of a quantum system in a particular desired state, when the system is subject to a projective measurement at independent identically distributed random time intervals. We present formulas for the probability of detection in the $n$th attempt.  We calculate as well  the mean and mean square both of the number of the first successful detection attempt and the time till first detection.  We present explicit results for a particle initially localized at a site on a ring of size $L$, probed at some arbitrary given site, in the case when the detection intervals are distributed exponentially.  We prove that, for all interval distributions and finite-dimensional Hamiltonians, the mean detection time is equal to the mean attempt number times the mean time interval between attempts. We further prove that for the return problem when the initial and target state are identical, the total detection probability is unity and the mean  attempts till detection is an integer, which is the size of the  Hilbert space (symmetrized about the target state). We study an interpolation between the fixed time interval case to an exponential distribution of time intervals  via the Gamma distribution with constant mean and varying width. The mean arrival time as a function of the mean interval changes qualitatively as we tune the inter-arrival time distribution  from very narrow (delta peaked) to exponential, as resonances are wiped out by the randomness of the sampling. 
 \end{abstract}
 
 \maketitle
 
 To read out the state of a quantum system, e.g., a quantum computer, some measurement must be performed.  Since one does not typically know when the computation will complete, monitoring of the system over time must be performed.  Being a quantum system, this monitoring comes at a price, since it interferes with the dynamics of the computation~\cite{Krapivsky2014,Li2018,Skinner2019,Nahum2020}.  One scheme of monitoring is that of repeated strong measurements at discrete time intervals.  In particular, the case of periodic probing, the so-called stroboscopic protocol, has received much recent attention~\cite{Bach2004,Krovi2006a,Krovi2006,Grunbaum2013,Bourgain2014,Dhar2015a,Dhar2015b,Sinkovicz2015a,Kuklinski2016,Friedman2017a,Friedman2017b,Thiel2018a,Thiel2018b,Nitsche2018a,Kuklinski2018,Yin2019,Meidan2019,Thiel2020,Thiel2020a,Liu2020,Kuklinski2020}. Here we focus on the problem of repeated probing at random, rather than constant, intervals~\cite{Varbanov2008}.  This includes, as a special case, that of almost periodic probing, which is of interest since in practice the intervals between measurements cannot be held exactly constant, and will have some degree of inherent noise.  We study the statistics of how many measurements it takes till the first successful detection of a given quantum state of the system, and how much time elapses till then, especially the low-order moments of these quantities.  We are particularly interested in how the results of this protocol compare to the statistics of the stroboscopic protocol.  It is known that stroboscopic measurements have certain advantages and disadvantages with respect to classical first-passage processes. One is that, even for small systems, the detection probability can be less than unity, due to so-called dark states, which is undesirable.  Other states, however, are detected very quickly due to constructive interference and this can lead to more efficient detection. Yet near a set of exceptional $\tau$s, the detection is highly inefficient, with the detection time diverging as these exceptional $\tau$ are approached. The ultimate question is how will the randomness of the time intervals between measurement help or hinder the detection process.
 
 It should be noted that this ``first detection" problem is a natural quantum analogue of the classic first-passage problem~\cite{Redner2001,Benichou2011,Metzler2014}.  In fact, for periodic probing, the generating function of the first-detection amplitude satisfies a renewal equation exactly parallel to that of the classic first-passage problem~\cite{Grunbaum2013,Friedman2017a}.
 This renewal equation relates the generating function of the first-detection amplitude/probability  to that the generating function of the free (i.e., unprobed) propagator. Introducing random monitoring times in the classical problem simply changes the free propagator.  Here, we will see that in the quantum case things are slightly more complicated due to the need to average over squared amplitudes, as opposed to probabilities.
 
 To set the stage, we will first review the basic principles of the stroboscopic protocol.  One prepares the system in a specified initial state $|\psi_\textrm{in}\rangle$.  One lets the system (living in an $N$-dimensional Hilbert space, $N$ finite) propagate unitarily under its Hamiltonian $H$ for some fixed time $\tau$, and then performs a strong measurement to detect whether the system is in the desired final state $|\psi_d\rangle$.  If the answer is yes, one stops there, but if the answer is no, one waits a further time $\tau$ during which the system continues its unitary evolution and performs another measurement, and so on until success. By the fundamentals of quantum mechanics, the unsuccessful measurements change the wave function, projecting out the $|\psi_d\rangle$ component, so that the combined propagation/measurement procedure breaks unitarity.  The probability of successful detection in the $n$th measurement is given by~\cite{Krovi2006a,Dhar2015a,Friedman2017a}
 \begin{equation}
 F_n = |\langle \psi_d | \phi(n)\rangle|^2
 \label{eq:Fn}
 \end{equation}
 where
 \begin{equation}
| \phi(n) \rangle \equiv  \left[U(\tau) \,P\right]^{n-1} U(\tau) |\psi_\textrm{in}\rangle
\label{eq:psin}
\end{equation}
is, up to a normalization factor, the wave function at time $n\tau^-$, right before the $n$th measurement, and $P\equiv {\cal I} -|\psi_d\rangle\langle \psi_d |$ projects out the detected state, ${\cal I}$ being the identity operator.  An illustrative example, which we will investigate in some depth, is the case when $H$ is a tight-binding Hamiltonian on a graph, say a ring of $L$ sites (so that here $N=L$), describing the hopping of a particle from site to site. We can prepare the system with the particle localized at a given site $x_\textrm{in}$ and ask when we first detect it at some (generally different) site $x_d$, probing the site $x_d$ after every interval $\tau$, with the system propagating freely between measurements, until the particle is successfully detected.  The statistics of this first successful detection, which arise from the interplay of the propagation and measurement, are the focus of our study.

 We now move beyond the stroboscopic protocol, with its fixed $\tau$, to the case where each successive $\tau$ is drawn from some fixed distribution.  At first glance, calculating the statistics for this seems a tall order, since for anything more complicated than a two level system it is well-nigh impossible to write down an explicit closed form expression for the $F_n$ as a function of the sequence $\tau_i$, $i=1,\ldots,n$, much less to average over this large number of random variables.
 Simulating this process is simple, at least in principle.  The simplest in principle is a direct simulation, starting the system at $\psi_\textrm{in}$, propagating for a random time interval drawn from the given distribution, measuring, and if undetected, propagating for another random interval and again measuring until successful, then repeating the whole process many, many times. A more efficient approach is to generate a sequence of IID $\tau$s, with probability density function $\rho(\tau)$ and then to implement the natural modification of Eq.  \eqref{eq:psin},
 \begin{equation}
| \phi(n) \rangle =  \left[\prod_{k=2}^{n} [U(\tau_n) P]\right] U(\tau_1) |\psi_\textrm{in}\rangle, 
 \label{eq:psinR}
 \end{equation}
 where the product is time-ordered from right-to-left in increasing $k$,
 and again  $F_n = |\langle \psi_d | \phi(n)\rangle|^2$.
   Then, one has simply to repeat this process a huge number of times with different sequences of $\tau$s, averaging to produce $\tauav{F_n}$, where the $\langle\cdot\rangle$ denotes an average over the $\tau$s:
\begin{equation}
\tauav{ F_n } = \iint \prod_i \left[\rho(\tau_i) d\tau_i \right] F_n(\{\tau_i\})
\end{equation}
 
\begin{figure}
  \includegraphics[width=0.45\textwidth]{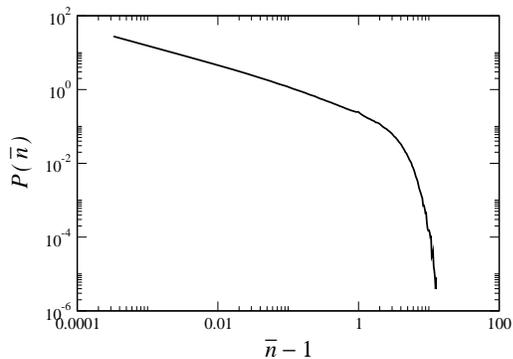}
  \caption{The distribution $P(\bar{n})$ of the mean measurements to detection  for a given realization of the inter-measurement intervals $\{\tau_k\}$, $\bar{n}$, constructed from  $10^6$ different realizations  for the return problem in the two-level system with hopping strength $\gamma=1$.  The $\tau$s were drawn from an exponential distribution with mean $\tauav{ \tau}=0.6$.  Note the square-root divergence near the lower limit of $\bar{n}=1$, due to cases where $\tau_1\ll 1$ and so ${\bar{n}}-1 \sim \tau_1^2$.  The distribution falls off exponentially at large $\bar{n}$.}
  \label{figbarnTLS}
  \end{figure}

 For the two level system, with Hamiltionian $H=-\gamma\left(|0\rangle\langle 1 | + |1\rangle\langle 0|\right)$, it is possible to express $F_n$ analytically. For example, for the so-called return problem, where $|\psi_d\rangle = |\psi_\textrm{in}\rangle = |0\rangle$,
 \begin{align}
 F_1 &= \cos^2 \gamma \tau_1; \nonumber\\
 F_n &= \sin^2 \gamma\tau_1 \left(\prod_{k=2}^{n-1} \cos^2 \gamma\tau_k \right)\sin^2 \gamma\tau_n
 \label{eq:FnTLS}
 \end{align}
 For each realization of the $\{\tau_k\}$, one finds that the total detection probability $P_\textit{det}\equiv \sum_n F_n =1$. However, each $\bar{n}\equiv \sum_n nF_n$ (as well as higher moments, of course), is different, even though for any fixed $\tau$ it is exactly 2. The distribution of $\bar{n}$ for $\gamma=1$ and exponentially distributed $\tau$s is displayed in Fig. \ref{figbarnTLS}. The measured mean of this distribution over $10^6$ realizations is $\tauav{ \bar{n}}=1.997$, in agreement with the analytic calculation of  2, derived in App. \ref{sec:TLS} directly from the $F_n$ in Eq.  \eqref{eq:FnTLS} above.  As we shall see below, the fact that $\tauav{\bar{n}}$ in the return problem is an integer, found by Gr\"unbaum, et al.~\cite{Grunbaum2013} for the stroboscopic case, is always true, independent of $H$ or the distribution of time intervals.  One can calculate analytically the variance of $\bar{n}$:
 \begin{equation}
 \textrm{Var}_\tau[\bar{n}] = \frac{2\, \textrm{Var}_\tau\left[\cos^2(\gamma\tau)\right]}{\left(1 - \tauav{ \cos^4(\gamma\tau)}\right)\left(1 - \tauav{ \cos^2(\gamma\tau)}\right )}
 \label{eq:Varbarn}
 \end{equation}
 where the averages and variance on the right-hand side are with respect to the single variable $\tau$, with measure $\rho(\tau)$, and which indeed is seen to vanish for the case of fixed $\tau$.  This evaluates to $ \textrm{Var}_\tau[\bar{n}] \approx 1.713$ for exponentially distributed $\tau$s with $\tauav{\tau}=0.6$, in excellent agreement with out direct simulation, which yielded $1.706$. This variance diverges for the stroboscopic protocol with $\tau$ an integer multiple of $\pi$, but is finite for any idstribution of $\tau$s with support outside these values.
  
 For a general Hamiltonian, the calculation of the $F_n$ and the averaging over the $\{\tau_k\}$ have to be carried out numerically and is not obviously amenable to analytic study.  Our first aim is to provide a solution for this problem.
We start with a reformulation of Eqs. \eqref{eq:psinR} and \eqref{eq:Fn}:
    \begin{equation}
    F_n = \Tr \left[ (\prod_{k=2}^n \hat{D}_k \hat{C} )\,\hat{D}_1 \hat{\Theta}\hat{B}\right]
    \label{eq:FnTr}
    \end{equation}
    The proof of this formula is in App. \ref{sec:AppFn}.  Here, $F_n$ is expressed in terms of a product of $N^2\times N^2$ matrices. We first define the $N\times N$ matrices
    \begin{align}
    (D_k)_{ij} &\equiv \delta_{ij} e^{-iE_i \tau_k}; \qquad B_{ij}\equiv 1; \qquad   \Pi_{ij} \equiv \delta_{ij} p_i \qquad  \nonumber\\
   C &\equiv {\cal I} - \Pi E; \qquad  \Theta_{ij }\equiv  \delta_{ij}\langle \psi_d| E_i \rangle \langle E_i |  \psi_\textrm{in} \rangle
    \end{align}
    where $|E_i\rangle$ is the $i$th energy eigenstate,  and $p_i \equiv |\langle E_i | \psi_d \rangle|^2$.
    Then, $N^2 \times N^2$ matrices are created via a Kronecker product~\cite{EOM}, via the hat notation $\hat{A}\equiv A^*\otimes A$ (see App. \ref{sec:AppFn} for definitions and details).

    This formula is similar to one obtained by Varbonev, et al.~\cite{Varbanov2008} who then specialized to the case of exponentially distributed $\tau$s.
      We can, however, formally take the average over the $\tau_k$s for an arbitrary distribution.  Since $\tau_k$ only appears in the factor $\hat{D}_k$ and the $\tau$s are independent, to take the average we simply need to replace each $\hat{D}_k$ in the product by its average $\tauav{ \hat{D}}$, which is independent of $k$.  This average is a diagonal matrix whose entries are just the characteristic function of the distribution of the waiting times, evaluated at the pairwise difference of energies:
    \begin{align}
    (\tauav{\hat{D}})_{(jk)(lm)} &= \delta_{jl} \delta_{km} \tauav{ e^{i(E_j - E_k)\tau} }\nonumber\\
     &= \delta_{jl} \delta_{km} \int_0^\infty   e^{i(E_j - E_k)\tau}\rho(\tau)d\tau
        \end{align}
    In particular, the ``diagonal" elements $(\tauav{\hat{D}})_{(jj)(jj)} = 1$. Thus, introducing ${\cal M} \equiv \tauav{ \hat{D} }\hat{C}$,
    \begin{align}
     \tauav{ F_n } 
     &= \Tr \left[{\cal M}^{(n-1)} \tauav{ \hat{D}} \hat{\Theta}\hat{B}\right]
     \label{eq:Fnavg}
     \end{align}
     
     This formula allows a direct calculation of the $\tauav{ F_n }$.  In particular, it works for the stroboscopic protocol of interval $\tau_0$ with $\rho(\tau) = \delta(\tau - \tau_0)$. It is  interesting to note that an immediate result is that $\tauav{ F_n }$ decays exponentially for large $n$ (accompanied by oscillations), at a complex rate determined by the eigenvalue of ${\cal M}$ closest to the unit circle.  It would naively appear that in general $\tauav{ F_n }$ is a linear combination of $N^2$ such decaying modes, which however does not square with what we know from the stroboscopic case.  There, the amplitude $|\phi(n)\rangle$ in Eq. \eqref{eq:psin} is a linear combination of $N-1$  dying (and, generally, oscillating) exponentials, corresponding to the $N-1$ nonzero eigenvalues $\lambda_i$ of the operator $U(\tau)P$ appearing in that equation. This implies that in this fixed $\tau$ case, $F_n$, the absolute square of the projection of $|\phi(n)\rangle$ on $|\psi_d\rangle$, is a linear combination of only $(N-1)^2$ modes, with decay rates $-\ln(\lambda_i + \lambda_j^*)$. The answer to this puzzle lies in the fact that, as discussed in App. \ref{App0Modes}, ${\cal M}$ has at least $2N-1$ zero modes, leaving at most $(N-1)^2$ nonzero eigenmodes.  
     
\begin{figure}
 \includegraphics[width=0.45\textwidth]{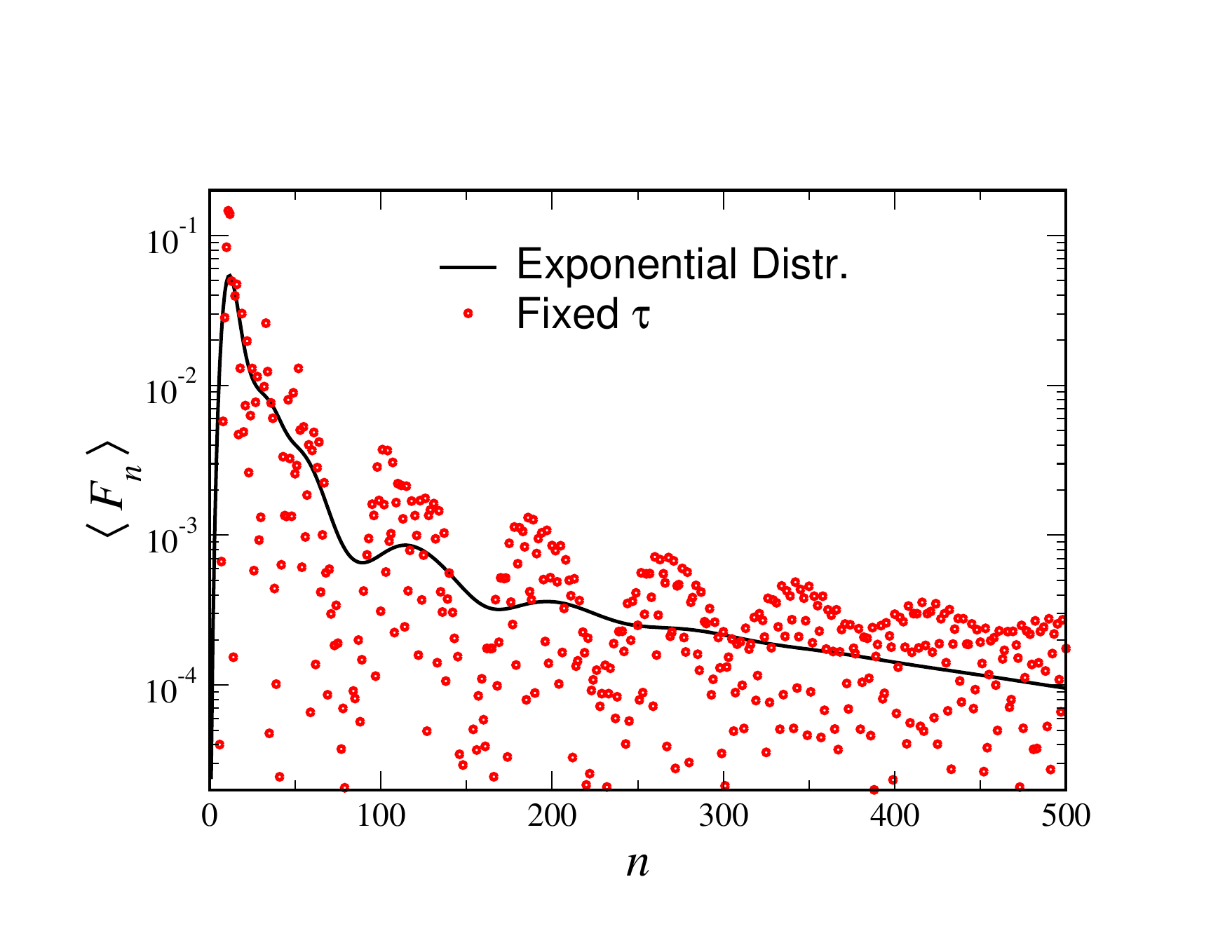}
 \caption{The distribution of the attempt number $n$ of first-successful detection for arrival at $x_d=12$ from an initial state localized at $x_\textrm{in}=0$ on a ring of length $L=24$. We compare the case of fixed $\tau=0.6$ with an exponentially distributed $\tau$, with mean $\tauav{\tau}=0.6$.  $\gamma=1$. The probabilities in both cases start near 0, rise rapidly until $n\approx 11-12$, and then decay exponentially in $n$ with superimposed oscillations. The oscillations for the fixed $\tau$ case are both larger, longer-lived, and more complex.}
 \label{fig:FnL24}
 \end{figure}

In Fig. \ref{fig:FnL24}, we show $\tauav{ F_n}$ for the arrival of the quantum walker at the detection site $x_d=0$, starting from the state localized at $x_\textrm{in}=12$, on a ring of size $L=24$,  with a nearest-neighbor hopping Hamiltionian $H = -\gamma \sum_k (|k\rangle\langle k-1| + |k\rangle\langle k+1|)$. We compare the case of a fixed $\tau=0.6$ with an exponential distributed $\tau$ with mean $\tauav{\tau}=0.6$.  The probabilities in both cases start near 0, rise rapidly until $n\approx 11-12$, corresponding to ballistic motion at the maximum group velocity~\cite{Thiel2018b}. Both then decay exponentially in $n$ with superimposed oscillations. The oscillations for the fixed $\tau$ case are both larger, longer-lived, and more complex, so that, as might be expected, the random $\tau$ case looks somewhat like a smeared-out version of the fixed $\tau$ case.  Also, the asymptotic decay rate of the magnitude is slower for the fixed $\tau$ case. For both cases, the total detection probability $P_\textrm{det}=\sum_n \tauav{ F_n}=1$. There is a difference in $\tauav{\bar{n}}$,  $\bar{n}$ averaged over the interval times, with $\tauav{\bar{n}} = 63$ for the exponential distribution and $\bar{n}\approx 101.4$ for fixed $\tau$.  This is an example where  the exponential sampling is much more efficient at detection than the stroboscopic protocol. 
     
%

  We can use our results to calculate $P_\textrm{det}$, $\tauav{\bar{n}}$, and $\tauav{\overline{n^2}}$ (and higher moments, if we desire) by summing the geometric series Eq.  \eqref{eq:Fnavg}.  We first define the matrix sum $S_F$:
  \begin{equation}
  S_F \equiv \sum_n [{\cal M}^{(n-1)} \tauav{ \hat{D}} \hat{\Theta}\hat{B}] = {\cal J}^{-1} \DD \hat{\Theta}\hat{B},
  \end{equation}
  where
\begin{equation}
{\cal J}\equiv {\cal I} - {\cal M},
\end{equation}
 and so
  \begin{align}
  P_\textrm{det} &= \Tr S_F = \Tr\left[{\cal J}^{-1} \DD \hat{\Theta}\hat{B}\right], \nonumber\\
  \tauav{\bar{n}} &\equiv \frac{1}{P_\textrm{det}}\sum_n n \tauav{ F_n}=  \frac{1}{P_\textrm{det}}\Tr \left[{\cal J}^{-2} \tauav{ \hat{D}} \hat{\Theta}\hat{B}\right], \nonumber\\
  \tauav{\overline{n^2}} &\equiv \frac{1}{P_\textrm{det}}\sum_n n^2 \tauav{ F_n}= \frac{1}{P_\textrm{det}}\Tr \left[{\cal J}^{-3} ({\cal I} + {\cal M}) \tauav{ \hat{D}} \hat{\Theta}\hat{B}\right].
  \label{eq:sums}
  \end{align}
  Here we have to insert a caveat - it can happen that ${\cal J}$ is singular~\cite{Krovi2006a}.  This happens precisely when there are dark states, energy eigenstates which are orthogonal to the detection state~\cite{Krovi2006,Thiel2020}.  This typically happens when the evolution operator $U(\tau)$ has a degeneracy in the spectrum. In such cases, one has to intepret Eq. \ref{eq:sums} in the sense of pseudo-inverses~\cite{Varbanov2008}.  Physically, this can be done by working
  from the very start in the restricted Hilbert space where the dark states have been eliminated, so that the dimension of  this new Hilbert space is $N_r\le N$.   In the problem of the ring with random $\tau$ or the fixed $\tau$ case with nonexceptional $\tau$, this is equivalent to working in the subspace of states symmetric about the detection site.  The detection state $|x_d\rangle$ can be expressed as a linear combination of these $N_r=\lfloor L/2+1\rfloor$ orthonormal states, which we can call $|\psi^e_i\rangle$, and so $\sum_i |\langle \psi^e_i | x_d\rangle|^2 =1$, whereas for $x_\textrm{in}\ne x_d, x_d + L/2$, $\sum_i |\langle \psi^e_i | x_\textrm{in}\rangle|^2 =1/2=P_\textrm{det}$.
In this last scenario, $\tauav{\bar{n}}$  is nominally infinite~\cite{Krovi2006a,Krovi2006,Varbanov2008}, but one can calculate instead a conditional average~\cite{Friedman2017a}, redefining $\tauav{\bar{n}}\equiv\langle \sum n F_n\rangle/P_\textit{det}$ (and similarly with other averages) so that it is the  expectation value, conditioned on successful detection, which is finite.  
 
\begin{figure}
 \includegraphics[width=0.45\textwidth]{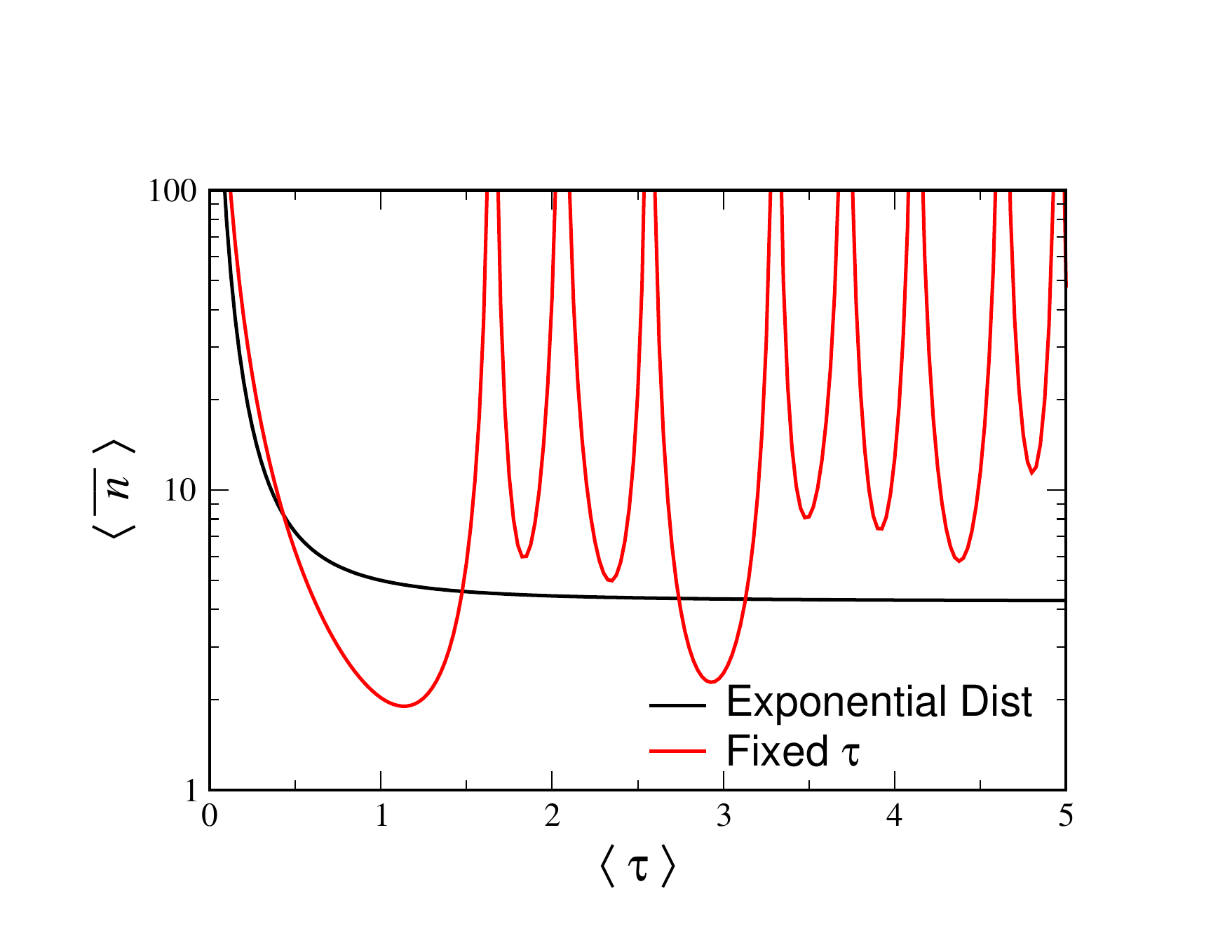}
 \caption{The mean of the attempt number, $\tauav{\bar{n}}$, of first-successful detection of arrival at $x_d=1$ from $x_\textrm{in}=0$ on a ring of length $L=7$. We compare the case of fixed $\tau=\tauav{\tau}$ with an exponentially distributed $\tau$, as a function of  $\tauav{\tau}$.  $\gamma=1$.}
 \label{fig:NbarL7Arr}
 \end{figure}

 In Fig. \ref{fig:NbarL7Arr}, we show $\tauav{\bar{n}}$ vs. $\tauav{\tau}$ for the  $0\to1$ arrival problem on a ring of size $L=7$, for both fixed $\tau$ and exponentially distributed $\tau$s.    For the arrival problem with fixed $\tau$,  $\bar{n}$ diverges  as $\tau\to 0$ as well as  each exceptional $\tau$ is approached~\cite{Friedman2017a,Friedman2017b}, this being a value of $\tau$ for which the operator $U(\tau)=e^{-iH\tau}$ has an accidental degeneracy due to the vanishing of $\tau(E_i -E_j) \mod 2\pi$ for some pair of energy eigenvalues $E_i, E_j$.  The divergence at small $\tau$ is a sign of the quantum Zeno effect~\cite{Thiel2020a}, and arises since in the limit $\tau\to 0$ all eigenvalues of $U(\tau)$ are unity, and hence degenerate.  For the arrival problem with exponentially distributed $\tau$'s,  $\tauav{\bar{n}}$  again diverges at small $\tauav{\tau}$, but  then falls monotonically with increasing $\tauav{\tau}$.  With fixed $\tau$ then, this implies that there is a value of $\tau$ which minimizes the conditional mean number of attempts to detection, whereas there is no such minimum with exponentially distributed $\tau$s.  In fact,  for a ring of size $L$ and exponentially distributed $\tau$'s, we can find $\tauav{\bar{n}}$ generally for the $0 \to x_d$ arrival problem:
 \begin{equation}
 \tauav{\bar{n}} = \left\{\begin{array}{cl} \frac{x_d(L-x_d)}{8\gamma^2\tauav{\tau}^2} + \frac{2L+3}{4},& L \textrm{ odd}, 1\le x_d < L \\[6pt]
                                                  \frac{x_d L}{8\gamma^2\tauav{\tau}^2} + \frac{L+3}{2} & L \textrm{ even}, 1\le x_d < L/2 \\[6pt]
                                                  \frac{L^2}{32\gamma^2\tauav{\tau}^2} + \frac{L+2}{2} & L \textrm{ even}, x_d = L/2 \end{array}\right.
                                                  \end{equation}
 These results were obtained by noticing that the analytic solution for $\tauav{\bar{n}}$ for small $L$ revealed the common structure $\tauav{\bar{n}}= A/\tauav{\tau}^2 + B$, with $A$ and $B$ rational.  This pattern was seen numerically to persist for larger $L$, and the coefficients $A$ and $B$ as functions of $L$ and $x_d$ were guessed and verified numerically for $3\le L \le 16$ and $1\le x_d \le L/2$.  Nevertheless, a proof for all $L$, $x_d$ remains to be achieved. The results below for $\tauav{\overline{n^2}}$ and $\tauav{\overline{t^2}}$ for the case of exponentially distributed $\tau$s on a ring were obtained similarly.  It is interesting to compare these results to the  classical first-passage time problem, either with fixed or random $\tau$. For small $\tau$, or equivalently $\tauav{\tau}$, the chance of crossing the absorbing site is very small, and the problem reduces to that of the first-passage to the end of an interval of length $L$, and so $\tauav{\bar{n}}$ is proportional to $x(L-x)/\tau$, as opposed to the $x(L-x)/\tauav{\tau}^2$ here, the difference being due to the quantum Zeno effect.   For large $\tau$, on the other hand, classically $\tauav{\bar{n}}$ goes to $L$, since the distribution after diffusing for a large time interval is constant, and so the probability of absorption in a single step is $1/L$ and the mean number of steps till absorption is $L$.
 
 The special nature of the case $L$ even, $x_d=L/2$ arises from the fact that here states even around $x_d$ are also even around $x_\textrm{in}=0$, and so $P_\textit{det}=1$ in this case, as noted above, whereas generally on a ring when $x_d\ne x_\textrm{in}$, $P_\textit{det}=1/2$.  This is reflected in the fact that 
$\lim_{x_d\to L/2} \tauav{\bar{n}} \sim L^2/(16\tauav{\tau}^2)$ for large $L$, whereas for $x_d=L/2$, $\tauav{\bar{n}} \sim L^2/(32\tauav{\tau}^2)$.
  
\begin{figure}
 \includegraphics[width=0.45\textwidth]{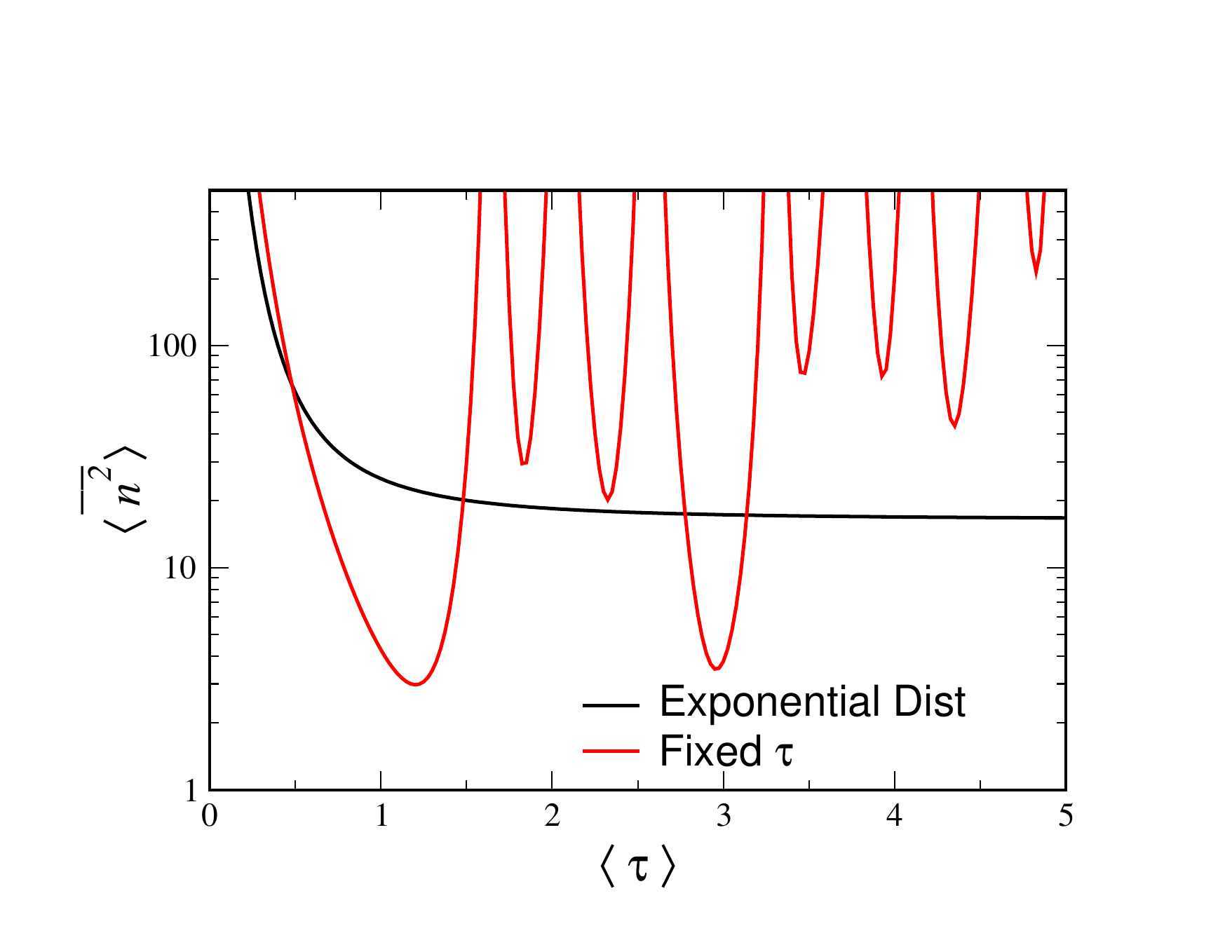}
 \caption{The mean squared  attempt number, $\tauav{\overline{n^2}}$, of first-successful detection at $x_d=0$, starting from $x_\textrm{in}=1$ on a ring of length $L=7$. We compare the case of fixed $\tau=\tauav{\tau}$ with an exponentially distributed $\tau$, as a function of $\tauav{\tau}$.  $\gamma=1$.  Notice the qualitative similarity with the corresponding curves for $\tauav{\bar{n}}$ in Fig. \ref{fig:NbarL7Arr}, as both are dominated by the divergences as $\tauav{\tau}\to 0$ (exponential distr. and fixed $\tau$) and at the exceptional $\tau$ (fixed $\tau$). The graphs of the ensemble variance $\tauav{ \overline{n^2}} - (\tauav{\bar{n}})^2$ (not shown) are also qualitatively similar.}
 \label{fig:N2barL7Arr}
 \end{figure}

 In Fig. \ref{fig:N2barL7Arr}, we show $\tauav{\overline{n^2}}$ for the $0\to 1$ arrival problem on the $L=7$ ring, again comparing the exponentially distributed $\tau$'s to fixed $\tau$, as a function of $\tauav{\tau}$. 
 The curves are quite similar to those for $\tauav{\bar{n}}$ shown in Fig. \ref{fig:NbarL7Arr}, and dominated by the divergences of the stroboscopic case at the exceptional $\tau$s, as well as the shared Zeno $\tauav{\tau}\to 0$ divergence. Here, the stroboscopic data show divergences for the return problem as well as the arrival problem, both at the exceptional $\tau$ as well as at $\tau=0$~\cite{Friedman2017a}. The exponential distribution data also diverges at $\tau\to 0$, with a $1/\tauav{\tau}^2$ divergence for the return problem and a stronger $1/\tauav{\tau}^4$ divergence for the arrival problem, and is again monotonically decreasing with increasing $\tauav{\tau}$. For this case, 
  \begin{widetext}

  \begin{equation}
 \tauav{\overline{n^2}} = \left\{\begin{array}{cl}  \frac{L(L+1)(L-1)}{48\gamma^2\tauav{\tau}^2} + \frac{2L^2+3L-1}{4} & L \textrm{ odd}, x_d=0 \\[4pt]
 \frac{Lx_d(L-x_d)(x_d(L-x_d) + 2)}{192\gamma^4\tauav{\tau}^4} + 
  \frac{L^3 + 2x_d(L-x_d) (L + 7) - L}{32\gamma^2\tauav{\tau}^2} + \frac{(4L^2+10L-3)}{8} &
 L \textrm{ odd}, 1\le x_d< L \\[4pt]
 \frac{L^3}{32\gamma^2\tauav{\tau}^2} + \frac{L(L+4)}{2} & L \textrm{ even}, x_d=0 \\[4pt]
 \frac{3x_d^2 L^3 - 4x_d(x_d^2-1)L^2}{384\gamma^4\tauav{\tau}^4} + 
 \frac{9L^3 + 12x_d L^2 + 24x_d(x_d+4)L
                                            - 16x_d(2x_d^2+1)}{192\gamma^2\tauav{\tau}^2} 
                  + \frac{(L^2+6L)}{2} & L \textrm{ even}, 1\le x_d < L/2 \\[4pt]
 \frac{L^3(L^2+8)}{3072\gamma^4\tauav{\tau}^4} + \frac{5L^3 + 12L^2 - 2L}{96\gamma^2\tauav{\tau}^2 }
                   + \frac{L^2 + 4L}{2} & L \textrm{ even}, x_d = L/2
  \end{array}\right.
 \end{equation}
 
 \end{widetext}
  The case where the detection state is identical to the initial state, the so-called return problem, exhibits some unique properties, as it does in the stroboscopic case~\cite{Grunbaum2013,Friedman2017a,Friedman2017b}.  Firstly, $P_\textit{det}=1$ for any distribution of $\tau$.  Secondly, $\tauav{ \bar{n}}$ is always an integer and independent of the distribution of the $\tau$s, and equal to  $N_r$, the dimensionality of the symmetrically reduced problem, as described above.  These statements are proved below.  For the fixed $\tau$ protocol, the accidental degeneracies at the exceptional $\tau$ imply that the reduced dimensionality and so also $\bar{n}$ are even lower at these special values.  Although the divergences in $\bar{n}$ are absent for the fixed $\tau$ return problem, they are present in the fixed $\tau$ results for $\overline{n^2}$.

We can also calculate $\tauav{\bar{t}}$, the conditioned mean time to detection. Obviously, for the stroboscopic protocol, $\bar{t}=\tau \bar{n}$.  This is not at all obvious for the case of random $\tau$s, since the $\tau$s and $n$s are coupled.  To compute $\tauav{\bar{t}}$, we need to compute $N_n \equiv \sum_n \sum_{k\le n} \tauav{ \tau_k F_n}$. For example,
\begin{align}
N_1&=\tauav{ \tau_1 F_1} = \tauav{ \tau\hat{D}} \hat{\Theta}\hat{B} , \nonumber\\
N_2&=\tauav{ (\tau_2+\tau_1) F_2} =  \tauav{ \tau\hat{D}} \hat{C} \tauav{\hat{D}} \hat{\Theta}\hat{B}  + \tauav{ \hat{D}} \hat{C} \tauav{ \tau\hat{D} } \hat{\Theta}\hat{B} ,\nonumber\\
N_n &= \tauav{ \tau\hat{D}}\hat{C} {\cal M}^{n-2} \tauav{\hat{D}} \hat{\Theta}\hat{B} + \tauav{ \hat{D} } \hat{C} N_{n-1}.
\end{align}
We can then sum the infinite series and take the trace to get
\begin{align}
S_N &\equiv \sum N_n =  {\cal J}^{-1}\left[ \tauav{ \tau\hat{D}} \hat{C} S_F +  N_1\right] \nonumber\\
\tauav{\bar{t}} &= \Tr S_N.
\end{align}
Evaluating this, we find that, even for random $\tau$'s, no matter their distribution, $\tauav{\bar{t}}=\tauav{\tau}\cdot \tauav{\bar{n}}$.  This is true both for the return and the arrival problem.

We can thus ask the question of which value of $\tauav{\tau}$ minimizes the mean detection time, conditioned on detection.  For the exponential distribution in the arrival problem on a ring, $\tauav{\bar{t}}$ takes the form $\tauav{\bar{t}} = A/\tauav{\tau} + B\tauav{\tau}$, $A$, $B$ being $L$-dependent coefficients, which is obviously minimized at ${\tauav{\tau}}_* = \sqrt{A/B}$.  Since for fixed $x_d$, both $A$ and $B$ grow linearly with $L$, ${\tauav{\tau}}_*$ approaches a constant value as $L \to \infty$ and so the optimal $\tauav{\bar{t}}$ grows linearly with $L$ as well.  When $x_d$ is of order $L$, then $A$ scales as $L^2$ while $B$ still scales as $L$, so that ${\tauav{\tau}}_*$ scales as $L^{1/2}$, and the optimal $\tauav{\bar{t}}$ scales as $L^{3/2}$.  All this is in sharp contrast to the stroboscopic protocol, where the $\tau$ which minimizes $\bar{t}=\tau \bar{n}$ is essentially controlled by the strong minimum that exists for $\bar{n}$ due to the divergences of $\bar{n}$ at $\tau \to 0$ and $\tau \approx \pi/2\gamma$.

\begin{widetext}
Proceeding similarly, we can calculate $\overline t^2$:
\begin{align}
\tauav{\overline{t^2}} &= \Tr \Bigg[{\cal J}^{-1}\Big(\tauav{ \tau^2\hat{D}} \hat{C} S_F + 2\tauav{ \tau\hat{D}} \hat{C} S_N  +  \tauav{ \tau^2\hat{D}} \hat{\Theta} \hat{B}\Big)\Bigg]
\end{align}
For the return problem, this is directly related to $\tauav{\overline{n^2}}$, namely $\tauav{\overline{t^2}} =( \tauav{\tau})^2 \tauav{\overline{n^2}} + N_r \textrm{Var}[\tau]$, for an arbitrary distribution.  There does not appear to be any such simple relation for the arrival problem.  Again, for exponentially distributed $\tau$s, we can calculate $\tauav{\overline{t^2}}$ explicitly:

 \begin{equation}
 \tauav{\overline{t^2}} = \left\{\begin{array}{cl}  
 \frac{Lx_d(L-x_d)(x_d(L-x_d) + 2)}{192\gamma^4\tauav{\tau}^2} + 
  \frac{L^3 + 2x_d(L-x_d) (L + 1) - L}{32} + \frac{(4L^2+14L+3)}{8} \tauav{\tau}^2&
 L \textrm{ odd}, 1\le x_d< L \\[6pt]
 \frac{3x_d^2 L^3 - 4x_d(x_d^2-1)L^2}{384\tauav{\tau}^2} + 
 \frac{9L^3 + 12x_d L^2 + 24x_d(x_d+1)L
                                            - 16x_d(2x_d^2+1)}{192} 
                  + \frac{(L^2+7L+3)}{2}\tauav{\tau}^2 & L \textrm{ even}, 1\le x_d < L/2 \\[6pt]
 \frac{L^3(L^2+8)}{3072\tauav{\tau}^2} + \frac{5L^3 + 3L^2 - 2L}{96 }
                   + \frac{L^2 + 5L+2}{2}\tauav{\tau}^2 & L \textrm{ even}, x_d = L/2
  \end{array}\right.
 \end{equation}
 
 \end{widetext}

lt just remains for us to prove our various claims above. In the following, we assume that we are working with the symmetry-reduced Hilbert space, so that the inverses are not singular and $N_r$ is the reduced dimension. We start with the observation that 
\begin{equation}
[\hat{B} {\cal J}^{-1}]_{(ij)(kl)} = \frac{1}{p_k} \delta_{k,l} \equiv W_{(ij)(kl)}
\end{equation}
with $p_k \equiv |\psi_k|^2$.
This is easily proved by multiplying out $W{\cal J}$ and recovering $\hat{B}$.  To evaluate the second term, one first notices that $W\tauav{\hat{D} } =W$, so only the ``diagonal'' columns of $W$ are nonzero, and the corresponding entries of $\DD$ are unity.  Then 
\begin{align}
[W\hat{C}]_{(ij)(kl)} &= \sum_{mn} \frac{1}{p_m} \delta_{mn}  (\delta_{mk}-p_m)(\delta_{nl}-p_n) \nonumber\\
&= \frac{1}{p_k}\delta_{kl} + \sum_m \frac{1}{p_m}\left(- p_m \delta_{km} -  p_m \delta_{ml} + p_m^2\right) \nonumber\\
&= \frac{1}{p_k}\delta_{kl} - 1 .
\end{align}
Thus, $W\hat{C}=W-\hat{B}$ and so $W{\cal J}=\hat{B}$, as desired.
From this follows the result that $\tauav{\bar{t}} = \tauav{\tau} \tauav{\bar{n}}$, since
\begin{align} 
\tauav{\bar{t}} &= \Tr  \left({\cal J}^{-1}\left[ \tauav{ \tau\hat{D}} \hat{C} S_F +  N_1\right]\right) \nonumber\\
&= \Tr \left({\cal J}^{-1}\left[ \tauav{ \tau\hat{D}} \hat{C}{\cal J}^{-1} \tauav{ \hat{D} }\hat{\Theta} \hat{B} + \tauav{ \tau \hat{D} } \hat{\Theta} \hat{B} \right]\right)\nonumber\\
&= \Tr \left(\hat{B}{\cal J}^{-1}\tauav{ \tau\hat{D}}\left[  \hat{C} {\cal J}^{-1} \tauav{ \hat{D} }\hat{\Theta}  + \hat{\Theta} \right]\right)\nonumber\\
&= \Tr \left(W  \tauav{ \tau\hat{D}} \left[ \hat{C}{\cal J}^{-1} \tauav{ \hat{D} }\hat{\Theta}  + \hat{\Theta}\right]\right)\nonumber\\
&= \tauav{\tau} \Tr \left(W \DD \left[  \hat{C}{\cal J}^{-1} \tauav{ \hat{D} }\hat{\Theta}  +  \hat{\Theta}\right]\right)\nonumber\\
&=\tauav{\tau} \tauav{\bar{n}}.
\end{align}
The key step here is that since only the ``diagonal" columns of $W$ are nonzero, and the corresponding elements of $\tauav{ \tau \hat{D}}$ are $\tauav{\tau}$, it follows that $\tauav{ \tau \hat{D}}$ can be replaced by $\tauav{\tau}\tauav{  \hat{D}}$.  

Our result about $W$ has another corollary regarding $P_\textrm{det}$:
\begin{align}
P_\textrm{det} &= \Tr \left[{\cal J}^{-1} \DD \hat{\Theta}\hat{B}\right] \nonumber\\
&= \Tr \left[\hat{B}{\cal J}^{-1} \DD \hat{\Theta}\right] \nonumber\\
&= \Tr \left[W \hat{\Theta}\right] \nonumber\\
&= \sum_i \frac{1}{p_i} p_i q_i  = \sum_i q_i
\end{align}
where $q_i = |\langle E_i | \psi_in\rangle|^2$.
For the return problem, $q_i = p_i$ and the sum of $p_i$ is unity, so $P_\textrm{det}=1$.  For the arrival problem, if there are no dark states, the sum of the $q_i$ is unity, whereas if there are dark states, then  if $|\psi_{in}\rangle$ has some overlap with a dark state, it follows that $P_\textrm{det}<1$.

We now turn to $\tauav{\bar{n}}$ in the return problem.  For the return problem, we note that $(S_F)_{(ij)(kl)} = \delta_{ij} p_i$.  We can prove this by computing ${\cal J}S_F = \DD \hat{\Pi} \hat{B}$, since in the return problem, $\hat{\Theta}=\hat{\Pi}$.  This computation is similar to the one for $W$ above, and involves the computation that $V\hat{B} - \hat{C} V \hat{B} = \hat{\Pi}\hat{B}$.  Given this,
we note that we can rewrite $S_F$ as $V\hat{B}$ where $V_{(ij)(kl)}=\delta_{ik}\delta_{jl}\delta_{ij} p_i$.
\begin{align}
\tauav{\bar{n}} &= \Tr \left[{\cal J}^{-1} S_F\right] \nonumber\\
&= \Tr \left[ E {\cal J}^{-1} V \right] \nonumber\\
&= \Tr \left[ W V\right] = \sum_i \frac{1}{p_i}p_i = N_r
\end{align}
It should be kept in mind that $N_r$ here is the dimension of the symmetric (i.e., non-dark) subspace.  If the distribution is discrete, it is possible that there are exceptional values of $\tauav{\tau}$
where this dimension is atypically small, and accordingly $\tauav{\bar{n}}$. For continuous distributions, this is not a concern. We see that the $\tau$ averaged value of $\bar{n}$ is quantised for any distribution of  waiting time PDF, including of course the $\delta$-function fixed $\tau$ distribution, thereby extending the stroboscobic measurement result of \cite{Grunbaum2013} concerning quantization of $\bar{n}$ in the return problem.

We next turn to $\tauav{\overline{t^2}}$ for the return problem.  We have
\begin{align}
\tauav{\overline{t^2}} &= \Tr \Bigg({\cal J}^{-1}\Big[ \tauav{ \tau^2\hat{D}} \hat{C} S_F +  \tauav{ \tau^2\hat{D}} \hat{\Pi} \hat{B} \nonumber\\
&\quad {}+2\tDD \hat{C} {\cal J}^{-1}\left[ \tDD \hat{C} S_F +  N_1\right] \Big]\Bigg) \nonumber\\
&=\Tr \Bigg(W\Big[ \tauav{ \tau^2\hat{D}} \hat{C} V +  \tauav{ \tau^2\hat{D}} \hat{\Pi} \nonumber\\
&\quad {}+ 2\tDD \hat{C} {\cal J}^{-1}\left[ \tDD \hat{C} V +  \tauav{\tau\hat{D}} \hat{\Pi} \right] \Big]\Bigg) \nonumber\\
&= \overline{\tau^2} \Tr \left(W\left[ \DD \hat{C} V +   \DD \hat{\Pi}\right]\right) \nonumber\\
&\quad {}+ 2\tauav{\tau} \Tr\left(W\DD \hat{C} {\cal J}^{-1}\left[ \tDD \hat{C} V +  \tauav{\tau\hat{D}} \hat{\Pi} \right] \right) \nonumber\\
\end{align}
The first two terms are relatively simple to treat, as their sum is simply $\overline{\tau^2} \tauav{\bar{n}}$. The last term, which we will label ${\cal T}$ requires more work:
\begin{align}
{\cal T} &= 2\tauav{\tau} \Tr \Big({\cal J}^{-1} \DD \hat{C} {\cal J}^{-1}\tDD \left[\hat{C} V\hat{B} +   \hat{\Pi}\hat{B} \right] \Big) \nonumber\\
&= 2\tauav{\tau} \Tr \left({\cal J}^{-1} \DD \hat{C} {\cal J}^{-1}\tDD\left[(V\hat{B} - \hat{\Pi}\hat{B}) + \hat{\Pi}\hat{B} \right]\right)\nonumber\\
&= 2\tauav{\tau} \Tr \left({\cal J}^{-1} \DD \hat{C} {\cal J}^{-1}\tDD V\hat{B} \right) \nonumber\\
&= 2(\tauav{\tau})^2 \Tr \left({\cal J}^{-1}\DD \hat{C} {\cal J}^{-1}\DD V\hat{B} \right) \nonumber\\
&= (\tauav{\tau})^2\left(\tauav{\overline{n^2}} - \tauav{\bar{n}}\right)
\end{align}
Thus, $\tauav{\overline{t^2}} = (\tauav{\tau})^2 \tauav{\overline{n^2}} + N (\overline{\tau^2} - (\tauav{\tau})^2)$, so that
$\tauav{\overline{t^2}} - (\tauav{\tau})^2 \tauav{\overline{n^2}} = N\textrm{Var}(\tau)$, our desired result.  Note that in the trivial case of $N=1$, when $n=1$ identically, this equation reads $\tauav{\overline{t^2}}= \tauav{\tau^2}$, which is obviously correct.

\begin{figure}
 \includegraphics[width=0.45\textwidth]{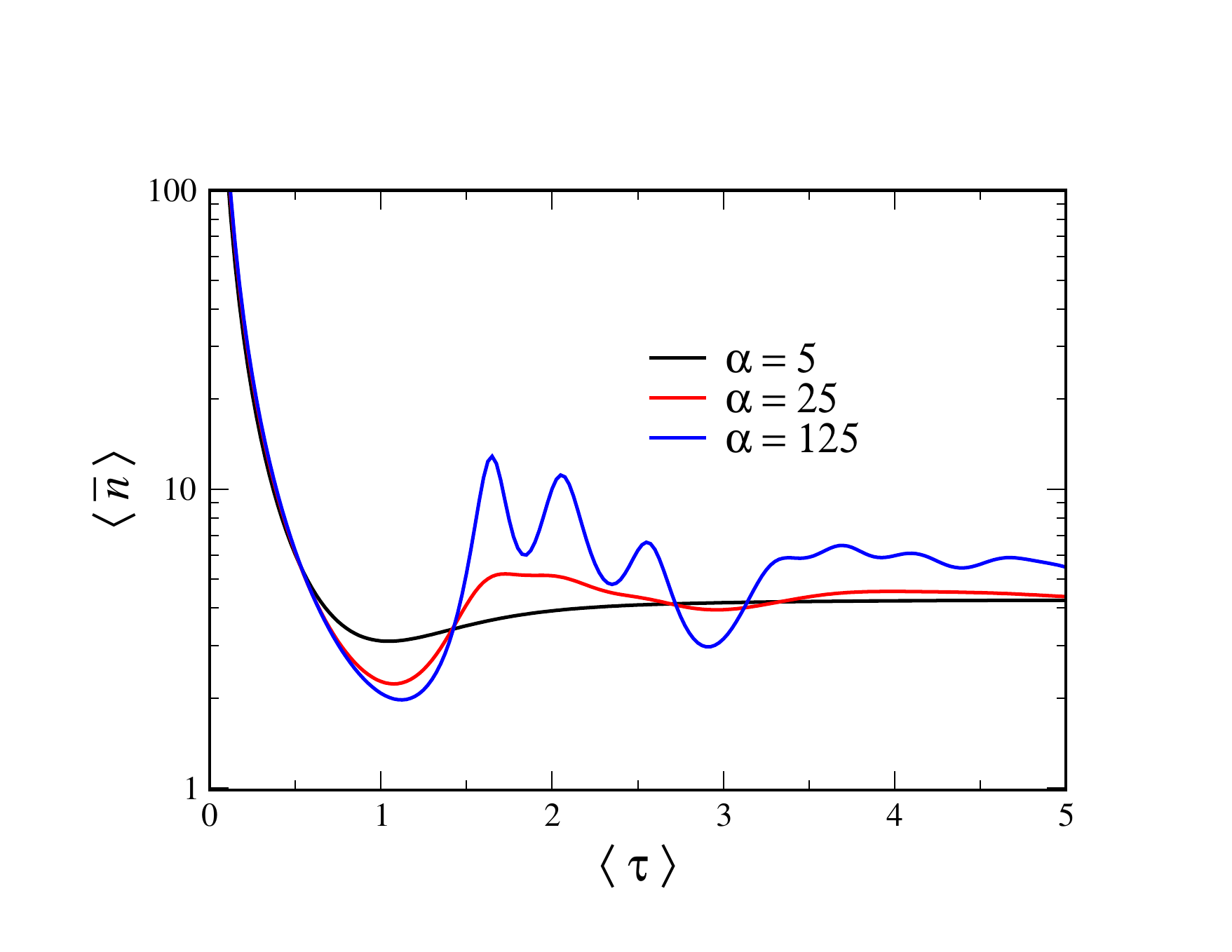}
 \caption{The mean  number, $\tauav{\bar{n}}$, of first-successful detection for the arrival at $x_d=0$ from $x_\textrm{in}=1$ on a ring of length $L=7$,  with a Gamma distribution of mean $\tauav{\tau}$ characterized by $\alpha=5,25,125$. Note that the extreme cases of an exponential distribution ($\alpha=1$) and the case of fixed $\tau=\tauav{\tau}$ ($\alpha\to\infty$) are presented in Fig. \ref{fig:NbarL7Arr}. For small $\tau$, all the various $\alpha$ yield similar results, while the oscillations increase in number and strength as $\alpha$ grows, turning into divergences in the infinite $\alpha$ limit. $\gamma=1$.}
 \label{fig:NbarL7GamArr}
 \end{figure}

Lastly, we turn to the question of what the effect of a little noise is on the stroboscopic results.  To examine this, we choose the Gamma distribution 
\begin{equation}
P_{\alpha,\beta}(\tau) = \frac{\beta^\alpha}{\Gamma(\alpha)}\tau^{\alpha-1} e^{-\beta\tau}
\end{equation}
with characteristic function $E[e^{is\tau}] = (1 - is/\beta)^{-\alpha}$.  As the mean of the distribution is $\alpha/\beta$, we fix $\beta=\alpha/\tauav{\tau} $.  Then, for $\alpha=1$ we have the exponential distribution, while in the limit $\alpha\gg 1$, we have a narrow distribution peaked at $\tauav{\tau}$ with the small variance $\tauav{\tau}^2/\alpha$.  We show in Fig. \ref{fig:NbarL7GamArr} $\tauav{\bar{n}}$ as a function of $\tauav{\tau}$ for $\alpha=5,25,125$.  We see the increase of $\alpha$ induces the oscillations characteristic of the stroboscopic data, with peaks whose height scales as the inverse of the squared  width of the distribution.  This is a consequence of the fact that $\tauav{ D} \sim  e^{-i\Delta \tauav{\tau}} + {\cal O}(1/\alpha)$ for large $\alpha$, where the first term  corresponds to the value of $\tauav{D}$ for the stroboscopic protocol.  This is consistent with our results for $\tauav{\bar{n}}$ for the arrival problem in the two level system, Eq. \eqref{TLSarv}, where
for a narrow distribution of width $\delta$ around $\tau=\pi/\gamma$, $1 - \tauav{\cos^2 \gamma \tau} \sim \delta^2$, leading to a $1/\delta^2$ divergence.

Thus, summing up, we have shown how to calculate the average over all sets of random independent $\tau$'s generated by some given distribution $\rho(\tau)$, both for the  probability distribution $F_n$ of the number of attempts till detection  and for its moments.  Using randomly generated $\tau$'s have been seen to preserve several of the salient features seen with the stroboscopic protocol. One major distinction is the presence of divergences and discontinuities associated with exceptional values of $\tau$ in the stroboscopic case, which do not exist for continuous distributions.  These divergences are ``tamed" into fluctuations, whose magnitude depend on the specifics of the $\tau$ distribution, $\rho(\tau)$. Among the universal features which survive unchanged is the nature of the Zeno, i.e., small $\tauav{\tau}$ limit. Also, various special features of the return problem, namely $P_\textrm{det}=1$ and $\tauav{\bar{n}}=N_r$ are preserved. A new property of the return problem, the simple relation between the mean squared number of attempts and the mean square time to detection is revealed. The introduction of random $\tau$s leads to a new set of quantities associated with the time till detection, which are correlated but not trivially related to the number of attempts till detection. Nevertheless, we have the general identity between the mean time and the average number times the mean interval between detections.  The correlations show up in the variance of the times.  One direction which is interesting to explore is to what extent we can relax the assumption of independent $\tau$s. We hope to report on this in the near future.  We are also investigating the connection of the quantization of the mean return time to Berry's phase and topology~\cite{Klaus2020}.
\begin{acknowledgments}
DAK and EB acknowledge the support of the Israel Science Foundation, 1898/17. KZ acknowledges the support of the Julian Schwinger Foundation. DAK acknowledges the assistance of Ruiqi Wang in the early stages of this work.
\end{acknowledgments}
       
    \appendix
    \section{Derivation of Eq. \eqref{eq:FnTr} \label{sec:AppFn}}
 We start by noting that the amplitude $\chi \equiv \langle \Psi | {\cal O}| \Phi \rangle$ for some operator ${\cal O}$  sandwiched between two arbitrary states can be written in the energy representation as
  \begin{equation}
  \chi = \sum_{k,l} \psi_k^* {\cal O}_{kl} \phi_l = \Tr [{\cal O} R E Q^*]
  \end{equation}
  where $\psi_k = \langle E_k| \Psi_k \rangle$, $\phi_k = \langle E_k| \Phi_k \rangle$, $E$ is the $N\times N$ matrix of all $1$'s, $R=\textit{diag}(\phi_1,\phi_2,\ldots,\phi_n)$, and $Q=\textit{diag}(\psi_1,\psi_2,\ldots,\psi_n)$.  With $|\Psi\rangle=|\psi_d\rangle$ and $|\Phi\rangle=|\psi_\textrm{in}\rangle$, and defining the matrices $\Theta \equiv Q^* R$, $D_j=\textit{diag}(e^{-iE_1 \tau_j}, e^{-iE_2 \tau_j},\ldots,e^{-iE_N \tau_j})$, we have, due to the cyclical property of the trace,
  \begin{equation}
  \langle \psi_d| \psi(1)\rangle = \Tr [D_1 REQ^* ] = \Tr [Q^* D_1 RE] = \Tr [D_1 \Theta E]
  \end{equation}
  since the diagonal matrices $D_1$ and  $Q^*$ commute.  Continuing, since $P = {\cal I} - QEQ^*$. we have
  \begin{align}
   \langle \psi_d| \psi(2)\rangle &=\Tr [D_2 ({\cal I} - QEQ^*) D_1 REQ^*] \nonumber\\
   &= \Tr [D_2 Q^*({\cal I} - QEQ^*) D_1 RE] \nonumber\\
   &=\Tr [D_2 ({\cal I} -\Pi E) Q^* D_1 QRE] \nonumber\\
   &= \Tr [D_2 C D_1 \Theta E]
   \end{align}
   where we have introduced $C\equiv {\cal I} -\Pi E$.  This generalizes to
   \begin{equation}
     \langle \psi_d| \psi^-(n)\rangle =\Tr [(\prod_{k=2}^n D_k C) D_1 \Theta E] .
   \end{equation}

   The point of this reformulation is it allows us to compute not only $\langle \psi_d| \psi^-(n)\rangle$, but its absolute square, i.e., $F_n$, through the use of the Kronecker product, since for any $N\times N$ matrices $A$, $B$,
   \begin{equation}
 \Tr A \cdot \Tr B=  \Tr [A \otimes B] 
    \end{equation}
   where the Kronecker product $(A\otimes B)$ is the $N^2 \otimes N^2$ block matrix, with the $i,j$ block, ($i,j$ as always running from $1$ to $N$), being the matrix $A_{ij} B$.  It will be convenient to label the elements of $A\otimes B$ by the compound indices $(j,k)(l,m)$ with $(A\otimes B)_{(j,k)(l,m)} = A_{jl}B_{km}$. Another useful identity is $(A_1\otimes B_1)(A_2\otimes B_2)=(A_1A_2)\otimes (B_1B_2)$.  Then,
   \begin{align}
   F_1 &= |\langle \psi_d| \psi(1)\rangle|^2 = \Tr [D_1^* \Theta E] \Tr [D_1 \Theta E] \nonumber\\
   &= \Tr [(D_1^*\otimes D_1)(\Theta^* \otimes \Theta)(E\otimes E)] \equiv \Tr [\hat{D}_1\hat{\Theta}\hat{B}]
   \end{align}
   where we have introduced the hat notation, $\hat{A} = A^* \otimes A$.
    Extending this to general $n$ gives us Eq. \eqref{eq:FnTr}.

\section{Two-Level System\label{sec:TLS}}
We work out here the results for the symmetric two-level system, first directly from Eqs. \eqref{eq:psin} and \eqref{eq:Fn}, and then using our general formulism.
What is special about the two-level system, with Hamiltonian $H=-\gamma(|0\rangle\langle 1| + |1\rangle\langle 0|)$ is that immediately after any measurement at the detector site $|0\rangle$, the particle is definitely in the state $|1\rangle$.  Using this, we have for the return problem $x_\textrm{in}=0$,
\begin{align}
|\psi(1)\rangle &= \cos \gamma \tau_1 |0\rangle + i \sin \gamma \tau_1 |1\rangle\nonumber\\
|\phi(n)\rangle &=A_n\left(\cos \gamma \tau_n |1\rangle + i \sin \gamma \tau_n |0\rangle\right)\nonumber\\
A_n &= i \sin \gamma\tau_1 \prod_{k=2}^{n-1} \cos \gamma\tau_k \nonumber\\
F_1 &= \cos^2 \gamma\tau_n \nonumber\\
F_n &= |A_n|^2\sin^2 \gamma\tau_n 
\end{align}
Thus, 
\begin{align}
\tauav{ F_1 } &=\tauav{ \cos^2 \gamma\tau }; \nonumber\\
\tauav{ F_n } &= \left[\tauav{ \sin^2 \gamma\tau } \right]^2 \left[\tauav{ \cos^2 \gamma\tau }\right]^{n-2}; \qquad n\ge 2 
\end{align}

We can recover these results from our new formalism.  We have
\begin{align}
\tauav{ \hat{D}} &= \textit{diag}(1, c + is, c-is, 1) \nonumber\\
\hat{\Pi} &= \hat{\Theta}= \textit{diag}(1/4,1/4,1/4,1/4) \nonumber\\
\hat{C} &= \frac{1}{4}\left(\begin{array}{cccc} 1 & -1 & -1 & 1 \\  -1 & 1 & 1 & -1 \\ -1 & 1 & 1 & -1 \\ 1 & -1 & -1 & 1\end{array}\right)
\end{align}
where $c\equiv \tauav{ \cos(2\gamma\tau)}$, $s\equiv \tauav{ \sin(2\gamma\tau)}$ and $\hat{B}$ is the $4\times4$ matrix of all $1$s. Then,
\begin{equation}
\tauav{ F_1 } = \Tr \frac{1}{4} \left(\begin{array}{cccc}
 1 & 1 & 1 & 1 \\   c+is& c+is&c+is&c+is\\
   c-is& c-is&c-is&c-is\\ 1 & 1 & 1 & 1\end{array}\right) = \frac{1+c}{2} = \tauav{ \cos^2 \gamma\tau }
\end{equation}
To calculate the other $F_n$, we need
\begin{equation}
{\cal M} = \frac{1}{4}\left(\begin{array}{cccc} 1&-1& -1& 1\\
  - (c+is)& c+is&c+is&-(c+is)\\
   - (c-is)& c-is&c-is&-(c-is)\\
 1&-1& -1& 1\end{array}\right)
 \end{equation}
 and, more generally, as may be proved by induction,
 \begin{equation}
{\cal M}^n = \left(\frac{1+c}{2}\right)^{(n-1)}{\cal M} \end{equation}
Then, for $n\ge 2$,
\begin{align}
\tauav{ F_n } &= \left(\frac{1+c}{2}\right)^{(n-2)} \times \nonumber\\
&\hspace{-0.25in}{} \Tr \left[\frac{1-c}{8} \left(\begin{array}{cccc}
 1 & 1 & 1 & 1 \\   -(c+is)& -(c+is)&-(c+is)&-(c+is)\\
   -(c-is)& -(c-is)&-(c-is)&-(c-is)\\ 1 & 1 & 1 & 1\end{array}\right)\right] \nonumber\\
   &= \left(\frac{1+c}{2}\right)^{(n-2)} \frac{(1-c)^2}{4} = (\tauav{ \cos^2 \gamma\tau })^{n-2)} (\tauav{\sin^2 \gamma\tau})^2
   \end{align}

From this, we get, with $C\equiv \tauav{ \cos^2 \gamma\tau }$
\begin{align}
P_\textit{det} &= C + \sum_{n=2}^\infty (1-C)^2 C^{n-2} = 1 \nonumber\\
\tauav{\bar{n}} &= C + \sum_{n=2}^{\infty} n(1-C)^2C^{n-2} = 2 \nonumber\\
\tauav{\overline{n^2}} &= C + \sum_{n=2}^{\infty} n^2(1-C)^2C^{n-2} = 2 + \frac{2}{1-C}
\end{align} 
We see immediately that $\tauav{\overline{n^2}}$ is finite as long as $C<1$.  For the fixed $\tau$ case, $C=\cos^2 \gamma\tau$, so $\overline{n^2}$ diverges for $\gamma\tau=k\pi$.  For any continuous distribution, $\overline{n^2}$ only diverges in the Zeno limit, in which case $C \to 1 - \gamma^2\tauav{\tau^2}$, so that $\tauav{\overline{n^2}} \sim 2/\gamma^2 \tauav{\tau^2}$. In comparing these results to those of the exponential distribution reported in the main text, it should be noted that $\gamma_{TLS}=2\gamma_{L=2}$ should be taken, since for our general ring, $E_k = 2\gamma\cos k$, while for the two level system, $E_k=\{-\gamma,\gamma\}$.

\begin{widetext}
We can similarly calculate $\tauav{\bar{t}}$, denoting $C_t \equiv \tauav{ t\cos^2 \gamma \tau}$
\begin{align}
\tauav{\bar{t}} &= C_t + \sum_{k=2}^\infty 2 \left[ (\tauav{\tau} - C_t) (1-C) C^{k-2} + (k-2)(1-C)^2 C_t C^{k-3} \right]\nonumber\\
&= C_t + 2(\tauav{\tau} - C_t)  + C_t = 2\tauav{\tau} ,
\end{align}
so that $\tauav{\bar{t}} = \tauav{\bar{n}}\tauav{\tau}$ as expected.
Also, defining $C_{tt} \equiv \tauav{ \tau^2 \cos^2 \gamma \tau}$,
\begin{align}
\tauav{\overline{t^2}} &= C_{tt} + \sum_{k=2}^\infty \Big[ 2(\tauav{\tau^2} - C_{tt})(1-C)C^{k-2} + (k-2) C_{tt}(1-C)^2 C^{k-3} \nonumber\\
&\qquad{} + 4(k-2)(\tauav{\bar{t}}-C_t)(1-C) C_t C^{k-3} \nonumber\\
&\qquad{} + (k-2)(k-3)(1-C)^2(C_t)^2 C^{k-4} + 2(\tauav{\tau} - C_t)^2 C^{k-2}\Big] \nonumber\\
&= C_{tt} + 2(\tauav{\tau^2} - C_{tt}) + C_{tt} + \frac{4(\tauav{\bar{t}}-C_t)C_t}{1-C} + \frac{2C_t^2}{1-C} + \frac{2(\tauav{\tau}-C_t)^2}{1-C} \nonumber\\
&= 2\tauav{\tau^2} + \frac{ 2\tauav{\tau}^2}{1-C} \nonumber\\
& = 2(\tauav{\tau^2} - \tauav{\tau}^2) + \tauav{\tau}^2 \tauav{\overline{n^2}}
\end{align}
again in line with expectation.
\end{widetext}

To calculate $\textrm{Var}(\bar{n})$, we need to calculate
\begin{equation}
\tauav{ (\bar{n})}= \sum_{k,l} kl \tauav{ F_k F_l }
\end{equation}
This calculation breaks up into three pieces. The first is the ``diagonal" contribution, $k=l$:
\begin{equation}
\sum_{k=1} \tauav{ k^2 F_k^2 } =C_4 +  \frac{(4 - 3 C_4 + C_4^2) S_4^2}{(1 - C_4)^3}
\end{equation}
where $C_4 \equiv \tauav{ \cos^4 \gamma \tau}$  and $S_4 \equiv\tauav{  \sin^4 \gamma \tau}$.  The contributions from $k=1$, $l>1$ and
$l=1$, $k>1$ are identical.
\begin{equation}
\sum_{k=1} \tauav{ k F_1 F_k } = \frac{(2 - C)  (C-C_4)}{(1 - C)}
\end{equation}
\begin{widetext}
Lastly, the contributions from $k\ge 2$, $l>k$ and $l\ge 2$, $k>l$ are also identical.
\begin{equation}
\sum_{k=2}\sum_{l=k+1} \tauav{ kl F_k F_l } = \frac{(6 - 4 C - 2 (3 - C_4) C_4 +  (3 - C_4)C C_4)  (C-C_4) S_4}{(1 -
   C) (1 - C_4)^3}
   \end{equation}
   \end{widetext}
Putting this  all together yields Eq. \eqref{eq:Varbarn} in the main text.

For the arrival problem, proceeding as above, we have
\begin{equation}
F_n = \sin^2 \gamma \tau_n \prod_{k=1}^{n-1} \cos^2 \gamma \tau_k
\end{equation}
In this case
\begin{align}
P_\textit{det} &=  \sum_{n=1}^\infty (1-C) C^{n-1} = 1 \nonumber\\
\tauav{\bar{n}} &= \sum_{n=1}^{\infty} n(1-C)C^{n-1} = \frac{1}{1-C} \nonumber\\
\tauav{\overline{n^2}} &=  \sum_{n=1}^{\infty} n^2(1-C)C^{n-1} =  \frac{1+C}{(1-C)^2}
\label{TLSarv}
\end{align}
and for the moments of the time
\begin{align}
\tauav{\bar{t}} &= \frac{\tauav{\tau}}{1-C} \nonumber \\
\tauav{\overline{t^2}} &= \frac{\tauav{ \tau^2}}{1 - C} + \frac{2\tauav{\tau}\tauav{\tau\cos^2 \gamma\tau}}{(1 - C)^2}
\end{align} 

\section{Zeros Modes of ${\cal M}$\label{App0Modes}}
Here we demonstrate that ${\cal M}$ has as least $2N-1$ zero modes. These eigenmodes arise from the fact that  $B$ has $N-1$ zero modes and one eigenvector with eigenvalue $N$, namely $(1;1;1;\ldots;1)$.  Thus, $\Pi B$ also has $N-1$ zero modes, with one eigenvector with eigenvalue $1$, namely $(p_1;p_2;\ldots;p_N)$.  In turn, $C={\cal I}-\Pi B$ has $N-1$ eigenvectors with eigenvalue unity, and one zero mode, $(p_1;p_2;\ldots;p_N)$.   Denote the $N-1$ zero eigenmodes as $|u_i\rangle$ and the remaining eigenvector with eigenvalue unity $|z\rangle$. Then, the vectors $|u_i\rangle \otimes |z\rangle$, $|z\rangle \otimes |u_i\rangle$ and $|z\rangle \otimes |z_\rangle$ constitute a set of $2N-1$ zero modes of $\hat{C}$ and hence of ${\cal M}$.   Thus, ${\cal M}$ has at most $(N-1)^2$ nonzero modes.  As long as all the elements of $\tauav{ \hat{D} }$ are nonvanishing,  as in the stroboscopic case as well for an exponential or gamma distribution of $\tau$, ${\cal M}$ has in fact exactly $(N-1)^2$ nonzero modes, consistent with our expectations.
 \bibliography{Article}

  \end{document}